\documentclass{llncs}
\usepackage{amsfonts,enumerate}
\newcommand{\A}{\mathbf{A}}
\newcommand{\B}{\mathbf{B}}

\renewcommand{\P}{\mathbf{P}}

\newcommand{\dom}{\mathrm{dom}\,}

\def\AlgGss{\mathrm{Alg}_{\Gamma_{ss}}}
\def\RNAlg{\mathrm{Alg}_{RNA}}

\def\NN{\mathbb{N}}

\begin{document}

\title{Modelling Biochemical Operations on RNA Secondary Structures\thanks{This work has been partially supported by the
Spanish DGES, grant BFM2000-1113-C02-01.}}

\author{Merc\`e Llabr\'es \and Francesc Rossell\'o}

\institute{Dept.\ of Mathematics and Computer Science,\\
Research Institute of Health Science  (IUNICS),\\
University of the Balearic Islands,\\
07122 Palma de Mallorca (Spain)\\
\email{$\mathtt{\{}$merce.llabres,cesc.rossello$\mathtt{\}}$@uib.es}}
\maketitle
\begin{abstract}
In this paper we model several simple biochemical operations on RNA 
molecules that modify their secondary structure by means of a suitable 
variation of Gro\ss e-Rhode's Algebra Transformation Systems.
\end{abstract} 
\section{Introduction}

Biochemical processes are responsible for most of the information 
processing that takes place inside the cell.  In the recent years, 
several representations and simulations of specific biochemical 
processes have been proposed using well known rewriting formalisms 
borrowed from Theoretical Computer Science.  Let us mention, for 
instance, Fontana's lambda calculus chemistry \cite{Font,FB}, recently 
revised by M\"uller \cite{Mul} (for a recent survey on artificial 
chemistry, see \cite{AC}), the stochastic Petri net approach 
\cite{GP}, the $\pi$-calculus representation of biochemical processes 
carried out by networks of proteins \cite{picalc}, and the graph 
replacement approach to DNA operations \cite{MN}.  In the latter, an 
\textsl{ad hoc} graph replacement formalism is developed to formalize 
DNA biochemical operations like annealing or denaturing, by 
considering DNA double strands to be special graphs.

There is another popular line of research in theoretical biochemistry 
that aims to represent the three-dimensional structure of biopolymers, 
and specially of DNA and RNA, by means of different kinds of formal 
grammars; see, for instance, \cite{May,Sear} for two surveys on this 
topic.  The ultimate goal of such a representation is to understand how 
the three-dimensional structure of a biopolymer is determined from its 
sequence of monomers (for instance, how the sequence of 
ribonucleotides of an RNA molecule determines its secondary structure; 
see below for the relevant details of RNA's biochemistry), and how 
this structure evolves when the biopolymer is modified through 
biochemical processes.

Sooner or later, this two lines of research should intersect, and the 
main goal of this paper is to move these two lines of research a step 
closer.  We formalize some simple biochemical processes on RNA 
molecules, like for instance ribonucleotide removals or mutations, and 
their effect on their three-dimensional structures by means of a 
variant of Gro\ss e Rhode's Algebra Transformation Systems (ATS) 
\cite{MGR} on partial algebras \cite{Bur86} representing RNA 
biomolecules.

Before entering into more details, it is time to introduce a little 
biochemistry.  As probably everybody knows, RNA molecules, together 
with DNA molecules and proteins, form the molecular basis of life.  An 
RNA molecule can be viewed as a chain of ribonucleotides, and each 
ribonucleotide is characterized by the base attached to it, which can 
be either adenine ($A$), cytosine ($C$), guanine ($G$) or uracil 
($U$).  An RNA molecule is uniquely determined by the sequence of 
bases along its chain, and it has a definite orientation.  Such an 
oriented chain of ribonucleotides is called the \emph{primary 
structure} of the RNA molecule.

In the cell and \emph{in vitro}, each RNA molecule folds into a 
specific three-dimensional structure that determines its biochemical 
activity.  To determine this structure from the primary structure of 
the molecule is one of the main open problems in computational 
biology, and partial solutions have been proposed using Stochastic 
Context Free Grammars and Dynamic Programming, among other tools; see, 
for instance, \cite[Chaps.\ 9, 10]{BSA}.

This three-dimensional structure is held together by weak interactions 
called \emph{hydrogen bonds} between pairs of non-consecutive 
bases.\footnote{Actually, for a hydrogen bond to be stable, the bases 
involved in it must be several nucleotides apart, but for simplicity 
we shall only consider the restriction that they must be 
non-consecutive.} Almost all these bonds form between 
\emph{complementary bases}, i.e., between $A$ and $U$ and between $C$ 
and $G$, but other pairings do also occur sporadically.  For 
simplicity, in this paper we shall only consider pairings between 
complementary bases.

In most representations of RNA molecules, the detailed description of 
their three-dimensional structure is overlooked and the attention is 
focused on its \emph{secondary structure}: the set of its base pairs, 
or \emph{contacts}.  Secondary structures are actually a simplified 
representation of RNA molecules' three-dimensional structure, but that 
is enough in some applications, as different levels of ``{graining}'' 
are suitable for different problems.
Two restrictions are usually added to the 
definition of secondary structure:
\begin{itemize}
\item If two bases $b_{i}$ and $b_{j}$ 
are paired, then neither $b_{i}$ or $b_{j}$ can bond with any other 
base; this restriction is called the \emph{unique bonds condition}. 

\item If  contacts exist between bases $b_{i}$ and $b_{j}$ and 
between bases $b_{k}$ and $b_{l}$, and if $b_{k}$ lies between $b_{i}$ and 
$b_{j}$, then $b_{l}$ also lies between $b_{i}$ and 
$b_{j}$; this restriction is called the \emph{no-pseudoknots condition}. 
\end{itemize}
The unique bonds condition simply captures the fact that the ``bond'' 
between two consecutive bases is of different nature, as a part of 
the molecule's backbone.  The 
no-pseudoknots condition is usually added in order to enable the use 
of dynamic programming methods to predict RNA secondary structures 
and, although real three-dimensional RNA structures have 
(pseudo)knots, we impose it here to show the scope of our approach: 
if pseudoknots are allowed, one simply has to allow them in the 
algebraic representation of RNA secondary structures and to remove the 
corresponding production rules from the rewriting system.
Thus,

This allows traditionally to represent an RNA molecule as a labelled 
graph, with nodes representing the ribonucleotides, and their labels 
denoting the bases attached to them, and arcs of two different kinds: 
ones representing the order of the bases in the primary structure (the 
\emph{backbone}) and the rest representing the bonds that form the 
secondary structure (the \emph{contacts}) \cite{RS96,SS99}.  Our 
representation is slightly different: the backbone is represented by a 
partial algebra corresponding, essentially, to a labelled finite 
chain, and then the contacts are specified as arcs of a graph on the 
nodes of the backbone.

There are some biochemical operations that can be carried out on an 
RNA molecule.  For instance, a ribonucleotide can be added or removed 
somewhere in the primary structure, a contact can form between two 
complementary bases, or it can be removed, and a base can mutate into 
another base.  These operations may have collateral effects: for 
instance, if a base mutates into another one and it was involved in a 
contact, then this contact will disappear, as the corresponding bases 
will no longer be complementary, and if a nucleotide is removed and as 
a consequence two nucleotides forming a contact become consecutive, 
then this contact will also break.

It is precisely when we tried to specify these side effects that we 
were not able to use simple graph transformation systems in an easy 
way, and we decided to use the ATS approach.  ATS is a very powerful 
algebra rewriting formalism, introduced by M. Gro\ss e-Rhode in 1999 
in order to specify the behavior of complex states software systems.  
It is operationally described, but not categorically formalized, and 
it takes care of side effects of the application of rules, similar to 
those found in our work.  Unfortunately, even the ATS formalism as 
defined in \cite{MGR}, which was designed with software engineering 
specification applications in mind, was not suitable, as it stands, 
for our purposes.  Thus we have slightly modified a simplified version 
of it, and we have dubbed the resulting formalism 
\emph{Withdrawal-based Algebra Transformation Systems} (WATS).  The 
reason is that, in our approach, the inconsistencies are eliminated by 
retreating, i.e., by removing, in a controlled way, the elements and 
operations that produce them, while in the original ATS approach the 
inconsistencies were eliminated by adding operations and identifying 
points.

The rest of this paper is organized as follows.  In Section 2 we 
represent RNA molecules as suitable partial algebras, in Section 3 we 
briefly introduce the WATS formalism, and then in Section 4 we show 
how to represent the aforementioned biochemical operations on RNA 
molecules by means of WATS production rules.  A final section on 
Conclusions closes the paper.

\section{RNA Molecules as Partial Algebras}

Roughly speaking, we represent the primary structure of an RNA 
molecule as a chain $\underline{n}$ of length $n\in \NN$ with a label 
in $\{A,C,G,U\}$ attached to each element of the chain, representing 
the base attached to the corresponding ribonucleotide, and its 
secondary structure by means of ordered pairs in $\underline{n}\times 
\underline{n}$.

Let $\Sigma_{ps}$ be the following many-sorted signature:
$$
\begin{array}{ll} 
 Sorts: & Nat, Bases \\
 Opns: & suc: Nat \to Nat \\
       & First, Last: \to Nat  \\
       & A, C, G, U: \to Bases \\
       & minor: Nat, Nat \to Nat \\
       & label: Nat \to Bases \\
       & \kappa : Bases \to  Bases
\end{array}
$$
An \emph{RNA primary structure} is a finite partial 
$\Sigma_{ps}$-algebra 
$$
\begin{array}{rl}
\P & =(P_{Nat},P_{Bases};\\ & \qquad First^{\P}, Last^{\P}, A^{\P}, 
C^{\P}, G^{\P}, U^{\P}, suc^{\P}, minor^{\P}, label^{\P}, \kappa^{\P})
\end{array}
$$
such that:
\begin{enumerate}[i)]

\item $(P_{Nat}; First^{\P}, Last^{\P}, suc^{\P})$ is a chain with 
successor operation $suc^{\P}$, first element $First^{\P}$ and last 
element $Last^{\P}$. 

\item The operation $minor^{\P}$ models the strict minority relation 
on this chain: $minor^{\P}(x,y)=x$ if and only if 
there exists some $n\geq 1$ such that $y=(suc^{\P})^n(x)$. 

\item The values of the nullary operations $A^{\P}, C^{\P}, G^{\P}, 
U^{\P}$ are pairwise different, $P_{Bases}=\{A^{\P}, C^{\P}, G^{\P}, 
U^{\P}\}$, and on this set the operation $\kappa^{\P}$ is given by the 
involution
$$
\kappa^{\P}(A^{\P})=U^{\P},\ \kappa^{\P}(U^{\P})=A^{\P},\ 
\kappa^{\P}(C^{\P})=G^{\P},\ \kappa^{\P}(G^{\P})=C^{\P}.
$$

\item The operation $label^{\P}$ is total.
\end{enumerate}
Notice that all these conditions except the last one cannot be 
specified through quasi-equations, since they are not satisfied by a 
trivial (with only one element of each sort) total 
$\Sigma_{ps}$-algebra.

Let $\Sigma_{ss}$ be now the signature containing $\Sigma_{ps}$ and, in 
addition, the following sorts and operation symbols:
$$
\begin{array}{ll} 
 Sorts:& Contacts \\
 Opns: & p_{1}:  Contacts \to  Nat\\
       & p_{2}: Contacts \to  Nat 
       \end{array}
$$

An \emph{RNA secondary structure} is a partial $\Sigma_{ss}$-algebra
$$
\begin{array}{rl}
\B=& (B_{Nat},B_{Bases},B_{Contacts};\\ & \qquad First^{\B}, Last^{\B}, A^{\B}, 
C^{\B}, G^{\B}, U^{\B}, suc^{\B}, minor^{\B}, 
label^{\B}, \kappa^{\B},p_{1}^{\B},p_{2}^{\B})
\end{array}
$$
whose $\Sigma_{ps}$-reduct is an RNA primary structure and it 
satisfies moreover the following quasi-equations:
$$
\begin{array}{ll} 
 (1) &  p_{1}^{\B} \mbox{ and } p_{2}^{\B} \mbox{ are total} \\[1ex]
  (2)    &  p_{1}^{\B}(x)=p_{1}^{\B}(y) \Rightarrow  x=y  \\[1ex] 
  (3)     &  p_{2}^{\B}(x)=p_{2}^{\B}(y)  \Rightarrow x=y \\[1ex] 
  (4)     &  p_{1}^{\B}(x)=p_{2}^{\B}(y)  \Rightarrow x=y \\[1ex] 
   (5)    &  minor(succ^{\B}(p_{1}^{\B}(x)),p_{2}^{\B}(x))=
   succ^{\B}(p_{1}^{\B}(x)) \\[1ex] 
   (6)    &  minor(p_{1}^{\B}(x),p_{1}^{\B}(y))=p_{1}^{\B}(x)\wedge
       minor(p_{1}^{\B}(y),p_{2}^{\B}(x))=p_{1}^{\B}(y) \\
       & \qquad\qquad\qquad\qquad
  \Longrightarrow minor(p_{2}^{\B}(y),p_{2}^{\B}(x))=p_{2}^{\B}(y) \\[1ex]
  (7)     & \kappa^{\B}(label^{\B}(p_{1}^{\B}(x)))=label^{\B}(p_{2}^{\B}(x)) 
 \end{array}
$$

In such an RNA secondary structure, each element $c$ of sort 
$Contacts$ represents, of course, a contact between nucleotides 
$p_{1}^{\B}(c)$ and $p_{2}^{\B}(c)$.  Equations (2), (3) and (4) 
represent the unique bonds condition, equation (5) represents the fact 
that there cannot exist a contact between a nucleotide and itself or 
its successor in the primary structure, equation (6) represents the 
no-pseudoknots condition, and equation (7) represents the fact that a 
contact can only pair complementary bases.
Notice that,  if we simply omit equation (6) then, pseudoknots are allowed
in the representation of RNA molecules.

Let $\Gamma_{ss}=(\Sigma_{ss},CE)$ be the specification whose set of 
consistence equations $CE$ are the quasi-equations (1) to (7) above.  
Let $\AlgGss$ the category whose objects are all partial 
\emph{$\Gamma_{ss}$-algebras}, i.e., those partial 
$\Sigma_{ss}$-algebras satisfying equation (1) to (7), and the 
morphisms between them are the plain homomorphisms, and let $\RNAlg$ 
be the full subcategory of $\AlgGss$ supported on the RNA secondary 
structures.

\section{Withdrawal-based $\Sigma_{ss}$-Algebra Transformation 
Systems}

Our Withdrawal-based Algebra Transformation Systems (WATS) are a 
modification of a simplified version of the Algebra Transformation 
Systems (ATS) introduced in \cite{MGR}.  This modification only 
affects the last step in the definition of the application of a 
rewriting rule through a matching, and therefore all definitions 
previous to that one are the same as in the original ATS formalism.  
Since we are only interested in rewriting RNA secondary structures, we 
shall only give the main definitions for the signature $\Sigma_{ss}$ 
introduced in the previous section.

So, to simplify the notations, let us denote by $S$, $\Omega$ and 
$\eta$ the set of sorts, the set of operation symbols and the arity 
function of the signature $\Sigma_{ss}$.  For every $\varphi\in 
\Omega$, set $\eta(\varphi)=(\omega(\varphi),\sigma(\varphi))\in 
S^{*}\times S$.

A $\Sigma_{ss}$-\emph{presentation} is a pair $P=(P_S,P_E)$ where
$P_S=(P_s)_{s\in S}$ is an $S$-set, whose elements will be called
\emph{generators}, and $P_{E}$ is a set of \emph{equations} with
variables in $P_{S}$
$$
t=t',\qquad t,t'\in  \mathrm{T}_{\Sigma_{ss}}(P_{S})_{s},\ s\in S.
$$
A special type of equations are the \emph{function entries}, of the
form 
$$
\varphi(\underline{a})=b, \qquad \varphi\in \Omega,\ \underline{a}\in
P_{S}^{\omega(\varphi)},\ b\in P_{\sigma(\varphi)}.
$$
A presentation is \emph{functional} when all its equations are 
function entries, and a functional presentation is \emph{consistently 
functional} when for every $\varphi\in \Omega$ and $\underline{a}\in 
P_{S}^{\omega(\varphi)}$, there is at most one function entry of the 
form $\varphi(\underline{a})=b$ in $P_{E}$.

Let $p:P_{S}\to P'_{S}$ be a mapping of $S$-sets.  If $e$ is an 
equation $t=t'$ with $t,t'\in \mathrm{T}_{\Sigma_{ss}}(P_{S})_{s}$, then we 
shall denote by $e[p]$ the equation $t(p)=t'(p)$ where $t(p),t'(p)\in 
\mathrm{T}_{\Sigma_{ss}}(P'_{S})_{s}$ are the terms obtained from $t$ and 
$t'$, respectively, by replacing all variables in them by their 
corresponding images under $p$.  In particular, if $e$ is the function 
entry $\varphi(\underline{a})=b$, then $e[p]$ stands for the function 
entry $\varphi(p(\underline{a}))=p_{\sigma(\varphi)}(b)$.  Given a 
mapping of $S$-sets $p:P_{s}\to P'_{s}$ and any set $E$ of equations 
with variables in $P_{S}$, let 
$$
E[p]=\{e[p]\mid e\in E\}.
$$  A 
\emph{morphism} of $\Sigma_{ss}$-presentations $p:(P_S,P_E) \to 
(P'_S,P'_E)$ is then a mapping of $S$-sets $p:P_S \to P'_S$ such that 
$P_{E}[p]\subseteq P_{E}'$.

A $\Sigma_{ss}$-\emph{rewriting rule} is a pair of $\Sigma_{ss}$-presentations, 
written $r =(P_l \leftarrow \rightarrow P_r)$, where $P_l=(X_l,E_l)$ 
and $P_r=(X_r,E_r)$ are functional presentations.  Informally, the 
left-hand side presentation in such a rule specifies the elements and 
operations that must be removed from the algebra which the rule is 
applied to, while its right-hand side presentation specifies the 
elements and operations to be added.  The generators that occur in a 
rule play the role of variables (and therefore we shall usually call 
them \emph{variables}): those appearing in the left-hand side 
presentation must be matched into the algebra to rewrite, and those 
appearing in the right-hand side presentation must be matched into the 
resulting algebra, in such a way that if a variable occurs in both 
parts of a rule, its image must be preserved.  In the sequel, we shall 
assume that all variables that occur in rewriting rules are taken from 
a universal $S$-set $X$ that is globally fixed and disjoint from all sets 
of operation symbols in the signatures we use.  We shall also assume 
that $X$ is large enough to contain equipotent copies of the carrier 
sets of all algebras we are interested in.

For every $\Sigma_{ss}$-rewriting rule $r=(P_l \leftarrow \rightarrow 
P_r)$, with $P_l=(X_l,E_l)$ and $P_r=(X_r,E_r)$, let:
 $$
\begin{array}{lll}
  X_l^0=X_l-X_r, &\qquad   & X_r^0=X_r-X_l, \\
  E_l^0=E_l-E_r, &\qquad   & E_r^0=E_r-E_l.
 \end{array}
 $$
For every $\Gamma_{ss}$-algebra $\A=(A,(\varphi^{\A})_{\varphi\in 
\Omega_{ss}}))$, let $A_{S}=A$ and 
$$
A_{e}=\{\varphi(\underline{a})=b \mid \varphi\in \Omega_{ss},\ 
\underline{a}\in \dom \varphi^{\A},\ \varphi^{\A}(\underline{a})=b\}.
$$
A \emph{match} $m$ for a $\Sigma_{ss}$-rewriting rule
$r=(P_l \leftarrow \rightarrow P_r)$ in $\A$ is simply a
presentation morphism $m:P_l \to (A_{S},A_E)$.
The \emph{extension} 
$$
m^*:X_l\cup X_r \to A_S\sqcup X_r^0
$$ 
of $m$ is defined by\footnote{As always, we identify any set with its 
image into its disjoint union with any other set.}
$$
m^*(x)=\left\{
\begin{array}{ll}
m(x)\in A_{S} & \mbox{if $x \in X_l$}\\
x  \in X_r^0 & \mbox{if $x \in X_r^0=X_{r}-X_{l}$}
\end{array}
\right.
$$

The \emph{application} of $r$ to $\A$ through $m$ rewrites then $\A$ 
into the partial $\Gamma_{ss}$-algebra $\B$ defined, step by step, as 
follows:

\begin{enumerate}[1)]

\item Set $B_{S}=(A_{S}-m(X_l^0))\sqcup X_r^0$.  This step removes 
from $\A$ the elements that are images of elements in $X_{l}$ that do no 
longer belong to $X_{r}$, and adds to it the elements in $X_{r}$ that 
did not belong to $X_{l}$.

\item Set $B_{E}=(A_E- E_l^0[m]) \cup E_r^0[m^*]$.  This step removes 
from $\A$ the operations that are images of function entries in 
$E_{l}$ that do no longer belong to $E_{r}$, and adds to it the 
equations in $E_{r}$ that did not belong to $E_{l}$, with variables in 
$B_{S}$.

\item Since the presentation $(B_{S},B_{E})$ is functional, it 
defines a partial $\Sigma_{ss}$-algebra with carrier set 
$B=B_{S}$ by simply translating the function entries in $B_{E}$ into 
operations; if this presentation is not consistently functional, then 
we must identify elements in $B$ in order to remove inconsistencies.  
This step can be formally described by means of a functor left adjoint 
to a functor that sends every $\Sigma_{ss}$-algebra to its 
presentation $(A_{S},A_{E})$.

\item If the $\Sigma_{ss}$-algebra defined in this way satisfies 
equations (1) to (7), we are done. Otherwise, there are two 
possibilities:
\begin{itemize}
\item Every contact $x\in B_{Contacts}$ that violates equations (1), 
(5) or (7) is removed.

\item After performing all  removals in the previous step, if 
there are still pairs of contacts $x,y\in B_{Contacts}$ that violate 
equations (2), (3), (4) or (6), then, if one of them comes from 
$X_r^0$ and the other comes from $A_{S}$, the one from $X_{r}^0$ 
is removed and the other one preserved, and otherwise both are 
removed.
\end{itemize}
\end{enumerate}

It is in step (4) where the main difference between Gro\ss e-Rhode's 
original ATS formalism and our WATS formalism lies.  In ATS, the 
$\Sigma_{ss}$-algebra obtained in (3) would be forced to satisfy 
equations (1) to (7) by taking its universal solution in $\AlgGss$, 
and thus adding operations and identifying elements.  In our 
formalism, violations of equations (1) to (7) are obviated by simply 
removing in a controlled way the contacts that yield them.

\section{Biochemical operations modelled by means of WATS}

The biochemical operations considered in this paper are the addition, 
deletion and mutation of a ribonucleotide and the addition and 
deletion of a contact.  Each of these biochemical operations can be 
modelled as a rewriting step of a WATS by the applications of a 
$\Sigma_{ss}$-rewriting rule to a RNA secondary structure.  The 
rewriting rules that model these biochemical operations are the 
following ones.

\begin{description}

\item[Adding a nucleotide:] We have to consider three different cases, 
corresponding to adding the new nucleotide at the beginning of the 
chain, at the end, or in the middle of it.  In this case, each rule 
must be understood as having a parameter $x$ which corresponds to the 
base attached to the new nucleotide.  So, there are four different 
values of this parameter, the nullary operation symbols $A$, $U$, $C$ 
and $G$.
\begin{itemize}

\item Rule $P_{add-base-first}(x)$ has: 
\begin{itemize}
\item as $P_l$ the set of 
variables $X_l=\{k_1\}$ of sort $Nat$ and the set of equations 
$E_l=\{First=k_1\}$;
\item as $P_r$ the set of variables 
$X_r=\{t,k_1,k_0\}$, of sorts $t \in Bases$ and $k_1,k_0\in Nat$, and 
the set of equations $E_r=\{First=k_0,suc(k_0)=k_1, 
label(k_0)=t,x=t\}$.
\end{itemize}

\item Rule $P_{add-base-last}(x)$ has
\begin{itemize}
\item as $P_l$ the set of variables $X_l=\{k_n\}$ of sort $Nat$ and 
the set of equations $E_l=\{Last=k_n\}$; 

\item as $P_r$ the set of 
variables $X_r=\{t,k_{n+1},k_n\}$, of sorts $t \in Bases$ and 
$k_{n+1},k_n\in Nat$, and the set of equations 
$E_r=\{Last=k_{n+1},suc(k_n)$ $=k_{n+1}, label(k_{n+1})=t, x=t\}$.
\end{itemize}

\item Rule $P_{add-base-middle}(x)$ has: 
\begin{itemize}
\item  as $P_l$ the set of 
variables $X_l=\{k_i,k_j\}$ of sort $Nat$ and the set of equations 
$E_l=\{suc(k_i)=k_j\}$;

\item as $P_r$ the set of variables $X_r=\{t,k_i,k_j,k\}$, of sorts $t 
\in Bases$ and $k_i,k_j,k \in Nat$, and the set of equations 
$E_r=\{suc(k_i)=k,suc(k)=k_j, label(k)=t,x=t\}$.
\end{itemize}
\end{itemize}

\item[Remove a nucleotide:] We have to consider again three different 
cases, corresponding to removing the nucleotide at the beginning of 
the chain, at the end, or in the middle of it.
\begin{itemize}

\item Rule $P_{del-base-first}$ has:
\begin{itemize}
\item as $P_l$ the set of variables $X_l=\{k_1,k_0\}$, both of sort 
$Nat$, and the set of equations $E_l=\{First=k_0,suc(k_0)=k_1\}$;

\item as $P_r$ the set of variables $X_r=\{k_1\}$ of sort $Nat$ and 
the set of equations $E_r=\{First=k_1\}$.
\end{itemize}

\item Rule $P_{del-base-last}$ has:
\begin{itemize}
\item as $P_l$ the set of variables $X_l=\{k_{n-1},k_n\}$, both of 
sort $Nat$, and the set of equations 
$E_l=\{Last=k_n,suc(k_{n-1})=k_n\}$;

\item as $P_r$ the set of variables $X_r=\{k_{n-1}\}$ of sort $Nat$ 
and the set of equations $E_r=\{Last=k_{n-1}\}$.
\end{itemize}

\item Rule $P_{del-base-middle}$ has:
\begin{itemize}
\item as $P_l$ the set of variables $X_l=\{k_i,k_j,k_l\}$, all of them 
of sort $ Nat$, and the set of equations 
$E_l=\{suc(k_i)=k_{l},suc(k_l)=k_j\}$;

\item as $P_r$ the set of variables $X_r=\{k_i,k_j\}$ of sort $Nat$ 
and the set of equations $E_r=\{suc(k_i)=k_j\}$.
\end{itemize}
\end{itemize}

\item[Mutating a base:] The mutation of a base is specified by just 
redefining the operation label.  Thus we consider the following rule:
\begin{itemize}
\item Rule $P_{mutation}$  has:
\begin{itemize}
\item as $P_l$ the set of variables
$X_l=\{x, y,k\}$, of sorts  $x,y \in Bases$ and $k\in Nat$, 
  and the set of equations $E_l=\{label(k)=x\}$;

\item as $P_r$ the set of variables $X_r=\{x,y,k\}$, of sorts $x,y \in 
Bases$ and $k\in Nat$ and the set of equations $E_r=\{label(k)=y\}$.
\end{itemize}
\end{itemize}

\item[Adding a contact:] To add a contact we simply add a new element 
of sort $Contact$ and the projections from it to the nucleotides it 
bonds.

\begin{itemize}

\item Rule $P_{add-contact}$ has:
\begin{itemize}
\item as $P_l$ the set of variables $X_l=\{x,y,k_i,k_{i+1},k_j\}$, of 
sorts $x,y \in Bases$ and $k_i,k_{i+1},k_j\in Nat$, and the set of 
equations $E_l=\{suc(k_i)$ $=k_{i+1}, minor(k_{i+1},k_j)=k_{i+1}, 
\kappa(x)=y,\kappa(y)=x, label(k_{i})=x,$ $label(k_{j})=y\}$;

\item as $P_r$ the set of variables $X_r=\{x,y,k_i, k_{i+1},k_j,c\}$, 
of sorts $x,y \in Bases$, $k_i,k_{i+1},k_j\in Nat$ and $ c\in 
Contacts$, and the set of equations $E_r=\{suc(k_i)=k_{i+1}, 
minor(k_{i+1},k_j)=k_{i+1}, \kappa(x)=y,\kappa(y)=x, p_1(c)=k_i, 
p_2(c)=k_j, label(k_{i})=x, label(k_{j})=y\}$.
\end{itemize}
\end{itemize}

\item[Remove a contact:] To remove a contact we simply delete it.

\begin{itemize}
\item Rule  $P_{del-contact}$ has:
\begin{itemize}
\item as $P_l$ the set of variables $X_l=\{k_i,k_j,c\}$, of sorts 
$k_i,k_j\in Nat$ and $c\in Contacts$, and the set of equations 
$E_l=\{p_1(c)=k_i, p_2(c)=k_j\}$;
 
\item as $P_r$ the set of variables $X_r=\{k_i,k_j\}$, both of sort 
$Nat$, and the set of equations $E_r=\emptyset$.
\end{itemize}
\end{itemize}
\end{description}

It is not difficult to check that an RNA secondary structure is always 
rewritten by the application of any one of these rules through any 
matching into an RNA secondary structure, and that in each case their 
effect is the desired one.  This must be done rule by rule and case by 
case.

\section{Conclusion}

We have modelled several simple biochemical operations on RNA 
molecules that modify their secondary structure by means of rewriting 
rules in a modified version of the Algebra Transformation Systems of 
Gro\ss e-Rhode, which we have dubbed Withdrawal-based Algebra 
Transformation Systems.  This modification has been made \textsl{ad 
hoc} for algebras representing RNA secondary structures, but we feel 
that the philosophy of removing inconsistencies by retreating 
should have applications in other contexts, and could probably be 
formalized for algebras over arbitrary specifications.

In this paper we have made some simplifications on the RNA secondary 
structure that could perfectly be avoided.  For instance, if we want 
to allow contacts between pairs of basis other than the usual 
complementary pairs, like for instance between G and U (they are 
called \textsl{wobble pairs}, not so uncommon), then we 
only have to replace the involution $\kappa$ by a symmetric relation 
on the carrier of sort $Bases$.  And if we want to impose that two 
bases paired by a contact must be at least at a fixed distance, we 
only have to modify in a suitable way equation (5).

There are also other collateral effects that could, and probably 
should, be specified.  For instance, isolated contacts tend to break, 
and pseudoknots should be allowed under certain circumstances.

\end{document}